# Spectral characterization of two-dimensional Thue-Morse quasicrystals realized with high resolution lithography

V Matarazzo,[1] S De Nicola,[1,*] G Zito,[1] P Mormile,[1] M Rippa,[1] G Abbate,[2] J Zhou,[3] L Petti[1]

[1]Istituto di Cibernetica "E. Caianiello" del Consiglio Nazionale delle Ricerche, Via Campi Flegrei 34, 80078 Pozzuoli (Na), Italy
[2]CNR-SPIN and Physics Dept., University of Naples "Federico II", I-80126 Naples, Italy
[3]Photonics Institute of the University of Ningbo City, Zhejiang Province, China

E-mail: s.denicola@cib.na.cnr.it

**Abstract**
One-dimensional Thue-Morse (ThMo) lattices are examples of self-similar structures that exhibit bandgap phenomena. ThMo multilayers may also possess fractal photonic bandgaps that give rise to large omnidirectional reflectance and light emission enhancement effects. Two-dimensional (2D) ThMo aperiodic quasicrystals possess interesting properties for photonic applications too. Here we demonstrate the experimental fabrication of large area 2D ThMo lattices into polymeric substrates at nanometer scale by electron beam lithography (EBL). Far field diffraction patterns of the experimental Thue-Morse structures have been measured and compared with the calculated theoretical Fourier spectra. Scanning electron microscopy and far field diffraction are used to characterize the experimental structures.

______________________________________________________________________

## 1. Introduction

In recent years, photonic quasicrystals (PhQCs) have attracted enormous interests in integrated optics and photonics. Quasiperiodic and aperiodic crystals may have high rotational symmetries not achievable by conventional periodic crystals [1, 2] and/or long-range translational symmetries that can induce the existence of large photonic bandgaps (PBGs) [3] with very interesting properties of light transmission [4], wave guiding and localization [5]. Such properties can be exploited in a wide variety of electro-optical and photonic applications. Moreover, the existence of high rotational symmetries and the presence of not equivalent defective states [6] opens the possibility to realize versatile, robust PBG devices even for low dielectric contrast materials like polymeric ones. Two-dimensional (2D) quasiperiodic lattices possessing high rotational symmetries have been largely studied in literature (see [7] and references therein).

One-dimensional (1D) aperiodic binary sequences based on specific substitution rules have been largely studied in literature too [8] and, recently, they have attracted an increasing interest for photonic applications [9]. The interplay between the optical properties and the underlying aperiodic lattice is a very interesting problem. An example of aperiodic substitutional lattice possessing a self-similar structure and exhibiting photonic bandgap phenomena is the Thue-Morse (ThMo) lattice [10-12]. 1D ThMo gratings are known to exhibit PGB with interesting omnidirectional reflectance [13]. In fact, in addition to the traditional photonic bandgap, the ThMo multilayers may exhibit the so-called fractal photonic bandgap, in case fractal structures are involved [14]. The scaling of the fractal bandgaps occurring in ThMo multilayers provides a large omnidirectional bandgap [15] and a significant light-emission enhancement effect for multiple resonance states [16].

Recently, theoretical studies on the 2D lattices based on substitutional ThMo sequences have shown the existence of photonic bandgaps too [17]. Experimental realization of photonic structures exhibiting 2D Thue-Morse arrangement has been reported only at the mesoscale range [18]. The experimental realization of two-dimensional PhQCs is a hard fabrication challenge. Two-beam and multiple-beam holographic lithography have been largely employed to realize periodic [19] and quasiperiodic [20] crystals at the mesoscale. $N$-fold symmetric aperiodically ordered quasicrystals are achievable through an accurate control of amplitude and phase of $N$ interfering laser beams. Different

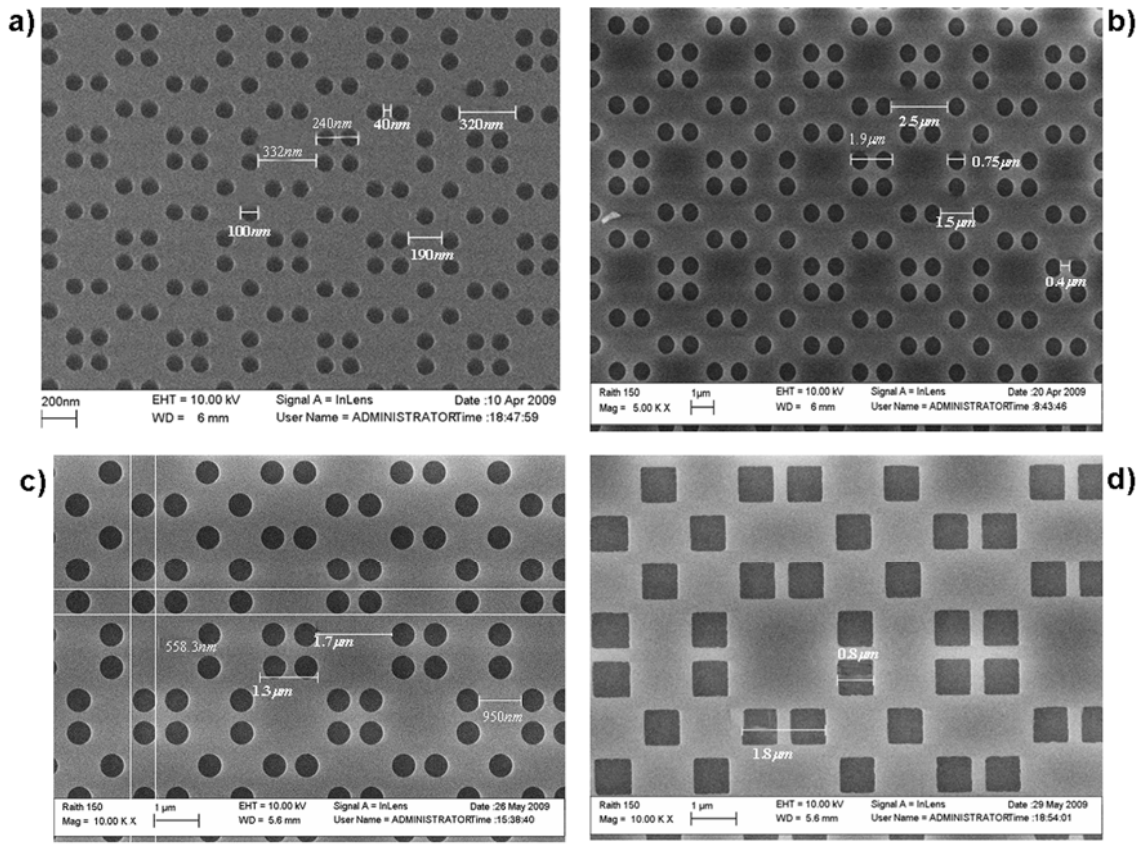

**Figure 1** (a)-(c) Scanning electron microscopy (SEM) images of 2D Thue-Morse quasicrystals obtained arranging circular air rods into a polymer substrate: (a) rod diameter $d$=100$nm$, lattice constant $a$=140$nm$; (b) $d$=750$nm$, $a$=1080$nm$; (c) $d$=500$nm$, $a$=720$nm$. (d) SEM image of a 2D ThMo structure obtained arranging square rods of side size $l$=800$nm$ from a regular periodic square lattice of constant $a$=1080$nm$.

geometries in the spatial tiling of the dielectric distribution of the medium deeply affect the overall PBG properties of the structures achievable [21]. Moreover, novel PBG aperiodic structures defined by recursive substitutional sequences like 1D and 2D Thue-Morse patterns cannot be realized even in principle by multiple-beam interference.

Differently, electron beam lithography (EBL) permits a very accurate control of the parameters of the structure to be realized with no limitation in the pattern design and tiling geometry allowing to realize 2D periodic, quasiperiodic and aperiodic crystals and allowing a very accurate control of the dielectric distribution of the exposed area so to produce feasible, high efficiency photonic bandgap structures.

In this work, the experimental fabrication of photonic quasicrystals into polymer by means of electron beam lithography is described. We experimentally realized, for the first time to our knowledge, several large-area 2D Thue-Morse aperiodic photonic lattices with nanometer resolution. The experimental structures that we realized have been characterized by scanning electron microscopy. Moreover, a spectral characterization of the ThMo samples has been provided by studying the far field diffraction patterns produced by the experimental structures. Experimental diffraction patterns have been compared with theoretical calculations of the Thue-Morse spectra. Furthermore, the analytical formulation of the two-dimensional ThMo spectrum was obtained. The spectrum can be related to the dispersion curves and hence to the photonic bandgap properties of the quasicrystal [22].

**2. Experimental realization of the Thue-Morse structures**

The fabricated experimental samples are 2D Thue–Morse quasicrystals. In the one-dimensional case, a binary Thue-Morse sequence is constructed using the following rule: given an arbitrary sequence of two symbols, $A$=0 and $B$=1, say, a new sequence is formed by replacing each occurrence of $A$ with the pair ($A$, $B$) and any occurrence of $B$ with the pair ($B$, $A$) [17]. The 1D ThMo grating is formed starting from the single element $A$. The structures are obtained by removing the lattice points from a square arrangement, following the inflation rules emerging from the Thue-Morse sequence. To realize an experimental dielectric structure for photonic application the elements $A$ and $B$ will correspond to particular regions of different dielectric materials. The sequence of order $N$ is then consisting of $2^N$ elements. By assuming $A=1$ and $B=-1$ (that is a $\pi$ phase jump between adjacent different dielectric elements), the intensity of the Fourier transform of the 1D ThMo sequence of order $N$ is given by

$$\left| f_N^{(1)}(k) \right|^2 = 2^{2N} \prod_{j=0}^{N-1} \sin^2\left(2^j \pi k\right), \qquad (1)$$

where $k$ is the magnitude of the reciprocal vector. Finite realizations of the ThMo lattice give rise to essentially discrete spectra (or diffraction patterns) giving evidence of the aperiodic nature of the crystal. By inflation rule the one-dimensional aperiodic lattice can be generalized to the two-dimensional space introducing a substitutional matrix obtained as Cartesian product of 1D ThMo sequences [17, 18].

In this work, the 2D ThMo structures have been experimentally obtained by removing particular lattice points from a regular square array of lattice constant $a$. The selected array elements are chosen in agreement with the inflation rules encoded into the substitutional matrix of the 2D ThMo sequence. Our experimental structures were fabricated by using a high resolution electron beam lithography (EBL) technique. The EBL facility employed consisted of a Raith 150 system. Such system enables the writing of patterns of arbitrary geometries with a spatial resolution up to 10$nm$. The samples were obtained by exposing a layer of positive photoresist (poly-methil-meta-acrylate PMMA) deposited on an indium tin oxide (ITO) coated glass. The e-beam is locally focused on the sample to expose selected regions of material homogenously along the depth of the substrate according to the calculated desired pattern. The resulting two-dimensional crystal is made of air rods of refractive index $n_a=1$ embedded into a polymeric matrix of PMMA possessing a refractive index $n_p \sim 1.5$. The centres of the air filled rods were located at the vertices of a 2D Thue-Morse lattice of order $N=10$. Scanning electron microscopy (SEM) images of several 2D ThMo quasicrystals experimentally realized are shown in figure 1. From figure 1-(a) to 1-(c), details of the ThMo patterns with cylindrical air rods are shown: (a) rod diameter $d=100$$nm$, lattice constant $a=140$$nm$; (b) $d=750$$nm$, $a=1080$$nm$; (c) $d=500$$nm$, $a=720$$nm$. In figure 1-(d) the SEM image of a 2D ThMo structure obtained arranging square rods of side size $l=800$$nm$ from a regular periodic square lattice of constant $a=1080$$nm$ is depicted. Large area structures, up to $800$$\mu m^2$, have been realized with nanometre resolution. The thickness of the samples was chosen in accordance to the characteristic sizes of the aperiodic structures. In order to test the versatility of the EBL technique to fabricate structures both at nanometer and submicron scales, several ThMo structures have been realized with different tiling parameters. The thickness of the substrate was in the range ~850-2000$nm$. With such large values of thickness it is difficult to create homogeneous holes with a regular size along the depth of the sample. The smallest value of rod diameter $d=100$$nm$ was chosen to test the maximum spatial resolution achievable with the EBL technique with such materials. The corresponding lattice constant $a=140$$nm$ was chosen from finite difference time domain (FDTD) simulations of the transmittance spectra of the ThMo structures in order to have a PBG in the visible. In figure 2, examples of the

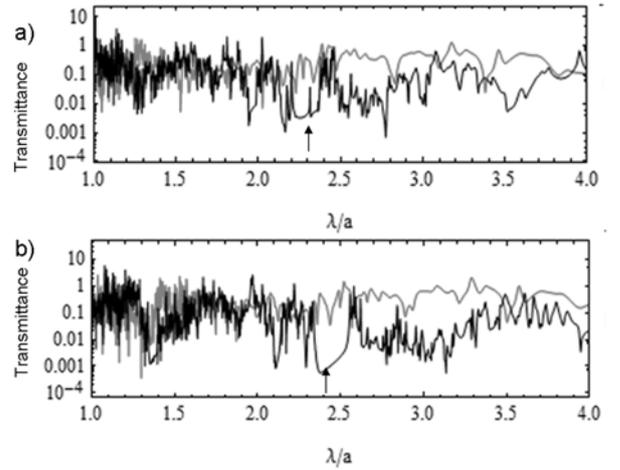

**Figure 2.** Transmittance spectra obtained from FDTD simulations of light propagation into ThMo structures of order $N=10$ for TM (grey curves) and TE (black curves) polarizations for refractive index difference $\Delta n=1.0$ (a) and $\Delta n=1.4$ (b).

transmittance spectra obtained with FDTD simulations [7] into 2D ThMo structures of order $N=10$ are shown as a function of the normalized wavelength $\lambda/a$. In figure 2, the bandgap is clearly evident in both insets (a) and (b) that correspond to a refractive index difference $\Delta n=1.0$ and $\Delta n=1.4$, respectively. Although the refractive index difference achievable with the PMMA is ~0.5, it is possible to use our samples as masks to realize photonic structures with higher dielectric contrasts or infiltrate the sample holes with a higher refractive material. However, further research is necessary in this direction.



The experimental characterization of the structures realized was particular difficult: the nanometre sizes of the lattice constant give rise to highly diverging diffraction orders difficult to collect due to the limited numerical apertures of the lenses used in our experimental setup. Hence, we realized ThMo structures with large values of rod diameters and corresponding lattice constants ($d$=750$nm$, $a$=1080$nm$; $d$=500$nm$, $a$=720$nm$) in order to acquire the experimental diffraction patterns of our samples. Moreover two kind of cell elements have been chosen, that is square and circular, because although standard photonic structures are made of circular rods, square based patterns can provide different properties that can be interesting. Furthermore, the experimental fabrication of square cells permitted to demonstrate the high potential of the technique in producing structures with uncommon pattern designs even at the nanometer scale.

The SEM characterization of the achieved experimental structures demonstrates the high spatial resolution that EBL technique allows in fabricating photonic quasicrystals.

## 3. Experimental diffraction patterns

The experimental Thue-Morse quasicrystals realized have been characterized in the direct space through SEM metrological measurements, whereas in the reciprocal space interesting properties of the aperiodic ThMo array can be provided through the experimental determination of the Fourier spectra of the samples. Generally speaking, the spectra can be related to the dispersion relations of the photonic structure providing important information on the bandgap structure of the quasicrystal [22]. The far field diffraction patterns of the experimental ThMo aperiodic samples have been acquired and compared with the expected theoretical spectra.

In figure 3, the experimental setup used to measure the Fourier spectrum of the Thue-Morse crystal is schematically depicted. The light source was a Ar$^+$ laser operating on the TEM$_{00}$ mode at the wavelength $\lambda$=514.5$nm$ and having a beam waist $w_0$=2$mm$. After having spatially filtered the laser beam through a lens-pinhole-lens system, the source is incident on the sample.

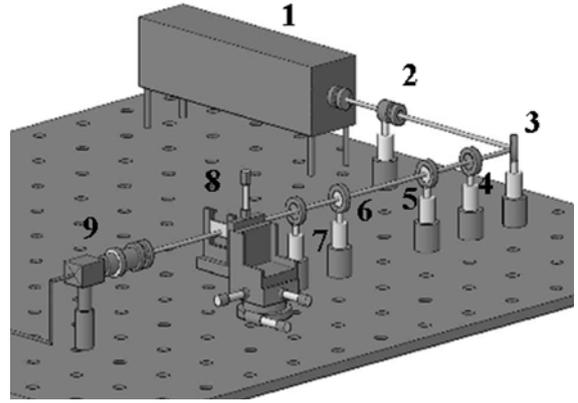

Figure 3. Schematization of the experimental set-up for spectra acquisition: 1) Laser Ar$^+$; 2) attenuator; 3) mirror; 4) pinhole of 500$\mu m$; 5) biconvex lens with focal length $f$=88.3$mm$; 6) biconvex lens with $f$=25.4$mm$; 7) pin hole; 8) sample positioning system; 9) CCD camera and focusing lens system.

Through the spatial filter, the incident light is diffracted creating an Airy pattern: the central zero order has an excellent spatial coherence and represents a good approximation of a plane wave. Higher diffraction orders have highly divergent wave vectors with respect to the optical axis. The second pinhole filters only the central spot of the Airy pattern. An optical system consisting of two confocal lenses is placed between the two pinholes providing a magnification ratio of 3.47 that ensures a uniform illumination of the experimental sample having a square surface of ~800$\mu m^2$. A CCD array coupled with a focusing lens system is finally used to acquire the Fourier spectra of the Thue-Morse structures.

The experimental diffraction pattern is then compared with the corresponding expected theoretical Fourier spectrum of the two-dimensional Thue-Morse structure. Starting from the one-dimensional formulation of the Fourier spectrum given in equation (1) for a Thue-Morse sequence of order $N$, we have generalized this result to the case of a two-dimensional aperiodic crystal. Let us indicate by $s$ the thickness of the sample and by $\Delta n$ the refractive index difference between different dielectric elements, that is between the air rods and the polymeric substrate of index $n_p$=1.5 and the air rods of index $n_a$=1. For a laser beam of wavelength $\lambda$, the optical phase mismatch $\phi$ experienced between adjacent elements is given by $\phi = 2\pi \Delta n s / \lambda$.



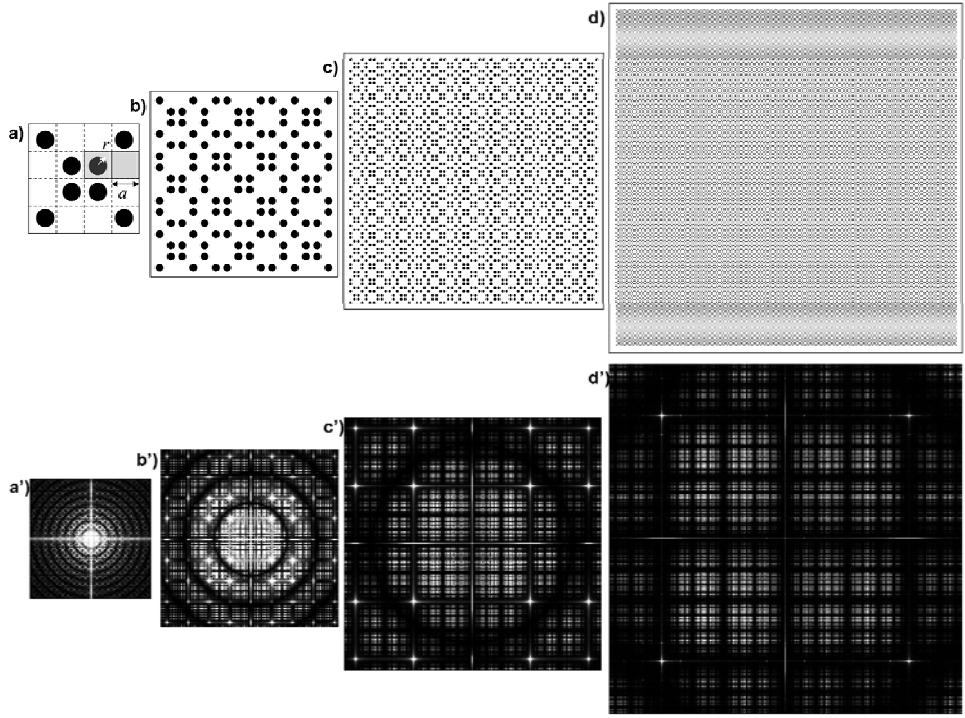

**Figure 4.** (a)-(d) Schematic representation of 2D Thue-Morse patterns of order $N$=2, 4, 6, 8, respectively; in inset (a) the schematization of the ThMo basic elements: top view of circular rod or radius $r$ inscribed into a square region of side size $a$. (a')-(d') Theoretical far field diffraction patterns of the corresponding ThMo lattices of order $N$=2, 4, 6, 8, respectively.

To calculate the diffraction pattern produced by a two-dimensional Thue-Morse structure, orthogonally oriented with respect to the impinging light, is convenient to separate the contributions of the dielectric circular rods of radius $r = d/2$ from the homogeneous environment of polymer in which the rod is embedded. If we consider a 2D Thue-Morse structure of order $N$, then the overall Fourier spectrum can be calculated considering that the whole pattern can be divided into $2^N \times 2^N$ equally sized square cells of two kinds as schematically represented in figure 4-(a) for the particular case $N$=2, that is one consisting of a circular rod into a square environment of different refractive index (grey cell on the left in figure 4-(a)), and one consisting of a homogenous square element (grey cell on the right of figure 4-(a)). The latter is a homogenous square dielectric region of side size $a$, hence the amplitude of the Fourier transform of such cell (corresponding to the polymer in the experimental sample), say $f_p(k_x, k_y)$, is simply given by

$$f_p(k_x, k_y) = \frac{\sin(\pi k_x a)\sin(\pi k_y a)}{\pi^2 k_x k_y}. \quad (2)$$

On the other hand, each region consisting of a circular rod of radius $r$ of index $n_r$ inscribed into a square cell of side size $a$ and index $n_p$ (left cell in figure 4-(a)), gives an overall contribution to the diffraction pattern, say $f_{r,p}(k_x, k_y)$, expressed by

$$f_{r,p}(k_x, k_y) = (1 - e^{i\phi}) r \frac{J_1\left[2\pi r \sqrt{k_x^2 + k_y^2}\right]}{\sqrt{k_x^2 + k_y^2}} + \quad (3)$$
$$+ e^{i\phi} f_p(k_x, k_y),$$

in which the first term containing the Bessel function of first order $J_1$ is associated to the circular rod, whereas the second term to the polymer environment of the same square cell. The two terms have a phase mismatch $\phi$. For the sake of simplicity let us indicate

$$u = e^{-2i\pi k_x a}, v = e^{-2i\pi k_y a} \quad (4)$$

as the basic Fourier spectral components. The two contributions to the ThMo spectrum of equations (2) and (3) must be properly combined in order to represent the diffraction pattern of the whole Thue-Morse structure. The total amplitude $f(k_x, k_y)$ of the Fourier spectrum can be written according to the following linear combination

$$f(k_x, k_y) = \alpha(k_x, k_y) f_{r,p}(k_x, k_y) + \quad (5)$$
$$+ \beta(k_x, k_y) f_p(k_x, k_y),$$



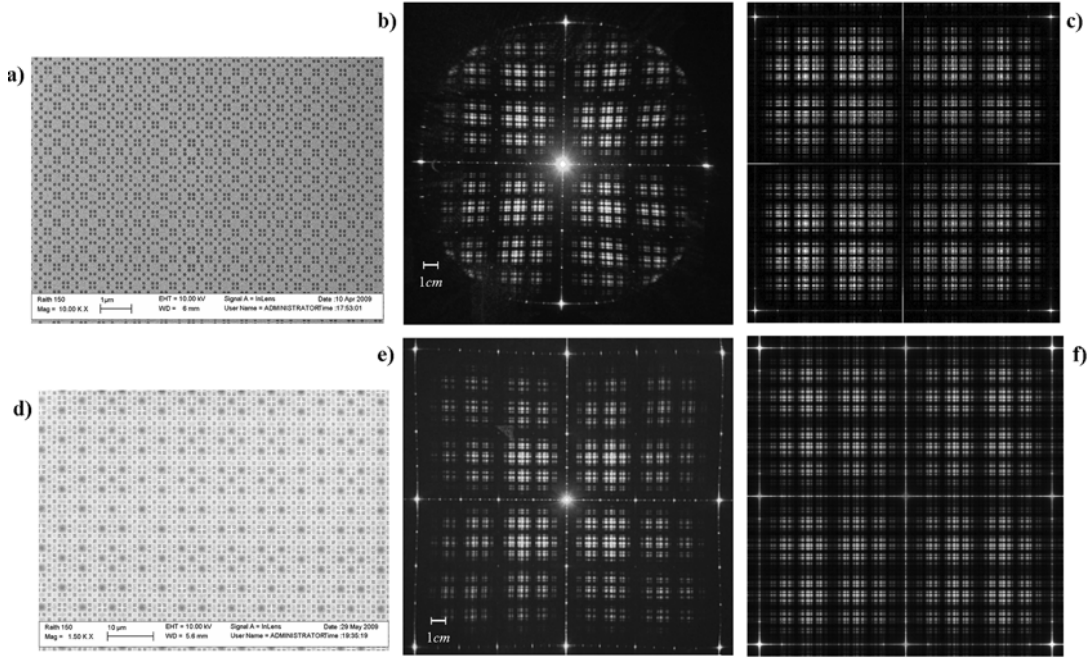

**Figure 5.** (a) Detail of an experimental 2D ThMo quasicrystal with circular air rods ($N=10$, $d=500nm$, $a=720nm$) ; (b)-(c) experimental and theoretical Fourier spectra, respectively, of the sample in (a). (d) Detail of an experimental 2D ThMo quasicrystal with square air rods ($N=10$, $l=800nm$, $a=1080nm$); (e)-(f) experimental and theoretical Fourier spectra, respectively, of the sample in (d).

where we have introduced the spatial frequencies dependent coefficients $\alpha(k_x,k_y)$ and $\beta(k_x,k_y)$. The coefficients $\alpha(k_x,k_y)$ and $\beta(k_x,k_y)$ are determined by taking into account that the coordinates of the centres of the circular rods are coincident with the spatial positions of the Thue-Morse lattice points. Following the ThMo inflation rule and using the spectrum of a 1D ThMo sequence given in equation (1), it is possible to verify that the coefficients of equation (5) for a 2D Thue-Morse quasicrystal of order $N$ can be written according to the following relations

$$\alpha(k_x,k_y) = \frac{uv}{2}\prod_{j=0}^{N-1}(1+u^{2^j})(1+v^{2^j}) + \\ + \frac{uv}{2}\prod_{j=0}^{N-1}(1-u^{2^j})(1-v^{2^j}), \quad (6)$$

$$\beta(k_x,k_y) = \frac{uv}{2}\prod_{j=0}^{N-1}(1+u^{2^j})(1+v^{2^j}) + \\ - \frac{uv}{2}\prod_{j=0}^{N-1}(1-u^{2^j})(1-v^{2^j}). \quad (7)$$

Thue-Morse quasicrystals of order $N=2, 4, 6, 8$, respectively, are schematically represented in figure 4-(a), 4-(b), 4-(c) and 4-(d), respectively. Figure 4 also displays the corresponding theoretical Fourier spectra calculated from the spectral distribution given by equation (5). It is clearly seen that the spectra becomes more structured with increasing the order of the Thue-Morse quasicrystal and it is difficult to recognize fine details even at relatively low orders. Examples of experimental diffraction patterns of Thue-Morse quasicrystals of order $N=10$ realized by e-beam lithography are shown in figure 5. Figure 5-(a) shows a detail of a Thue-Morse quasicrystal with circular air rods; the experimental and theoretical diffraction patterns, corresponding to the same sample are shown in (b) and (c), respectively. The agreement between the experimental and theoretical spectra is good. Figure 5-(d) shows a detail of a Thue-Morse quasicrystal with square air rods, whereas in (e) and (f) the corresponding experimental and theoretical diffraction patterns, respectively, are shown. The analytical spectrum is calculated in this case substituting equation (3) with the following expression

$$f_{r,p}(k_x,k_y) = (1-e^{i\phi})\frac{\sin(\pi k_x l)\sin(\pi k_y l)}{\pi^2 k_x k_y} + \\ + e^{i\phi}f_p(k_x,k_y), \quad (8)$$

in which the first term is related to the diffraction produced by a square of side size $l$ and refractive index $n_a$ embedded into a square cell of size $a$ and index $n_p$.

Distortions of the spectra shown in figures 5-(b) and (e) are caused by the finite numerical aperture of the collecting lens.

## 4. Conclusions

As in the case of one-dimensional Thue-Morse lattices, ThMo multilayers may possess very interesting photonic properties like fractal bandgaps that give rise to almost unknown optical phenomena. In this work, we have demonstrated the experimental fabrication of large area 2D ThMo lattices into polymeric substrates at nanometer scale by electron beam lithography. To our knowledge, this is the first experimental realization of sub-micrometric Thue-Morse quasicrystal structures. The quality of the pattern design has been analyzed by means of scanning electron microscopy confirming that EBL allows a very accurate control of the writing pattern with a spatial resolution up to 10 $nm$. Experimental results of the far field diffraction pattern have been compared with the analytical calculation of the Fourier spectra and have been shown to be in good agreement. By using the analytical calculation of the Thue-Morse spectra it is possible to calculate the photonic bandgap diagrams [22] of the Thue-Morse quasicrystals.

8